\magnification 1200
\headline{\hfil\tenrm\folio\hfil}
\footline={\hfil}
\headline{\ifnum\pageno=1\nopagenumbers
\else\hss\number\pageno \hss\fi}
\footline={\hfil}
\overfullrule=0pt
\baselineskip=20pt
\hfuzz=5pt \null

\def\R{ {\rm R \kern -.31cm I \kern .15cm}}

\input epsf
\vskip 1 cm
\centerline{{\bf DIFFRACTIVE DISSOCIATION IN DEEP}}
\centerline{{\bf INELASTIC SCATTERING AT HERA}}
\par
\bigskip
\centerline {by}
\medskip
\centerline{{\bf A. Capella, A. Kaidalov}\footnote{$^+$}{Permanent address :
ITEP, B.
Cheremushkinskaya 25, Moscow, Russia.}, {\bf C. Merino and J. Tran Thanh Van}}
\centerline{Laboratoire de Physique Th\'eorique et Hautes
Energies\footnote{*}{Laboratoire
associ\'e au Centre National de la Recherche
Scientifique - URA 63.}} \centerline{B\^atiment 211, Universit\'e
de Paris-Sud, 91405 Orsay cedex, France}
\vskip1cm

\noindent \underbar{\bf Abstract} \par
The diffraction dissociation of virtual photons is considered in the framework
of
conventional Regge theory. It is shown that the recent HERA data on large
rapidity gap events can
be successfully described in terms of the Pomeron structure function. Using
Regge factorization,
the latter can be related to the deuteron structure function. The parameters
which relate these two structure functions are determined from soft hadronic
diffraction data. The
size of the shadowing corrections at low $x$ and large $Q^2$ is also obtained.

\vskip2cm
 \noindent{\obeylines LPTHE Orsay 94-42
\noindent May 1994}
\vfill\eject
\baselineskip=20pt
Experiments at HERA provide new possibilities to study diffractive processes
and the properties of
the Pomeron. Recently, spectacular large rapidity gap events in deep inelastic
scattering at very small $x$
have been observed by Zeus$^{[1]}$ and H1$^{[2]}$ collaborations at HERA. On
the other
hand, diffractive production has been studied for many years in high energy
hadronic interactions
and a large amount of information on these reactions has been obtained (see for
example
reviews$^{[3-5]}$). A new unexplored region of diffraction production by highly
virtual photons is
now accessible at HERA. \par

Here we shall consider the process of diffractive dissociation of highly
virtual
photons in the framework of Reggeon theory, which is the basis for the
description of diffractive
processes in hadronic interactions. We shall demonstrate that the
cha\-rac\-te\-ris\-tic features of
the data on large rapidity gaps events can be understood using the Regge based
description of
structure functions, introduced in a previous work$^{[6]}$, together with our
knowledge of
diffractive production in high-energy soft hadronic collisions. \par

The amplitude of the single diffractive production of a hadronic state $X$ is
described by the
Pomeron exchange diagram shown in Fig. 1. The initial particle $i$ in this
figure can be either a
hadron or a photon (real or virtual). This diagram corresponds to single
diffractive
production without dissociation of the initial proton, while double diffraction
corresponds to
dissociation of both projectile and target. At high energies $s >> m^2$, when
the total rapidity
interval $\xi = \ell n {\displaystyle s \over\displaystyle{m^2}}$ is large,
these diagrams describe
kinematical configurations of final particles in which large rapidity gaps
$(\Delta y)$ are present.
\par

Consider first the single diffraction dissociation of a real particle $i$ (Fig.
1). In the pole
approximation for Pomeron exchange the cross section of this process is given
by (see
e.g.$^{[3]}$)

$$s_1 {d \sigma_D \over dt \ ds_1} = {1 \over 16 \pi} \left ( g_{pp}^P(t)
\right )^2 exp \left [ 2
\left ( \alpha_P(t) - 1 \right ) \xi ' \right ] \sigma_{iP}^{tot}(s_1, t)
\eqno(1)$$
\vskip 3 mm

\noindent where $s_1 \equiv M^2$ is the square of the invariant mass of the
hadronic system $X$, $t =
(p' - p)^2$ is the invariant momentum transfer to the final proton and
$g_{pp}^P(t)$ is the vertex
describing the Pomeron coupling to a proton. The Feynman-$x$ of the Pomeron is
defined by $x_P =
{\displaystyle{s_1} \over\displaystyle s}$, and $\xi ' = \ell n \ s/s_1 \approx
\Delta y$. The last
term in eq. (1) is the total Pomeron-particle cross-section$^{[7]}$. It is
obtained by summation
over all final states $X$ and over their phase space. Note that this quantity,
contrary to the cross
section of usual particles, is just defined by eq. (1), but cannot be
interpreted as a physical
observable on its own. At large $s_1 >> m_i^2$ this cross section is determined
by the Pomeron and
Regge exchanges and has the form (see Fig. 2)~:

$$\sigma_{iP}^{tot}(s_1, t) = \sum_k g_{ii}^k(0) \ r_{PP}^k(t) \left ( {s_1
\over s_0} \right
)^{\alpha_k(0)-1} \ \ \ , \eqno(2)$$
\vskip 3 mm

\noindent with $s_0 = 1 \ {\rm GeV}^2$. Here $\alpha_k$ denotes the trajectory
of any reggeon $k$
contributing to the $i$-$P$ elastic amplitude, and $g_{ii}^k(0)$ and
$r_{PP}^k(t)$ are the
couplings of the reggeon $r$ to particle $i$ and to the Pomeron, respectively
(see Fig. 2). The only
reggeons $k$ that contribute to eq. (2) are the Pomeron itself and the
$f$-trajectory. Note that in
the first factor of eq. (1) we have only considered the Pomeron contribution.
The contribution of
secondary reggeons such as the $f$-trajectory is important only for $x_P >
0.05$ to $0.1$. \par

In the case when particle $i$ is a virtual photon (with
virtuality $q^2 = - Q^2$), $\sigma_{\gamma^{\ast}P}^{tot}$ and
$g_{ii}^k$ in eqs. (1) and (2) depend on $Q^2$. For large values of $Q^2$, it
is
convenient to introduce the structure function of the Pomeron$^{[8-10]}$,
$F_P$, related to the
total cross-section of a virtual photon $\sigma_{\gamma^{\ast}P}^{tot}$ in the
same way as the total
cross section of a virtual photon on a proton is related to the proton
structure function $F_2$~:

$$\sigma_{\gamma^{\ast}P}^{tot}(s_1, Q^2, t) = {4 \pi^2 \alpha_{em} \over Q^2}
F_P(s_1, Q^2, t) \ \ \
. \eqno(3)$$
\vskip 3 mm

Note that in the present case there is an extra variable $t$ (the virtuality of
the Pomeron). It is
natural to assume$^{[8-10]}$ (and calculations of the simplest QCD diagrams
confirm
it$^{[10-15]}$) that at large $Q^2$ the function $F_P$ obeys approximate
Bjorken scaling (up to a
logarithmic dependence on $Q^2$, given by QCD-evolution), i.e. depends
essentially on the variable

$$x_1 = {Q^2 \over 2 \left ( p_{_P} \cdot p_i \right )} \simeq {Q^2 \over s_1 +
Q^2} \ \ \ .
\eqno(4)$$
\vskip 3 mm

It follows from eq. (4) that the characteristic masses in the diffractive
production by
highly virtual photon are $s_1 \sim Q^2$. For fixed values of $x_1$ the
$Q^2$-dependence of the
diffraction production process is the same $(1/Q^2)$ as for the total cross
section of the virtual
photon on the proton, i.e. it is a main twist effect. On the other hand the
diffractive production
of a state with fixed mass (e.g. $\rho$, $\omega$, $\phi$, ...) decreases
faster with $Q^2$. \par

In the case of virtual photons we also have the Bjorken variable

$$x = {Q^2 \over 2 \left ( p_p \cdot p_i \right )} \simeq {Q^2 \over s + Q^2} \
\ \ .$$
\vskip 3 mm

\noindent The longitudinal momentum fraction taken by the Pomeron is $x_P =
x/x_1$, and,
consequently, the variable $\xi '$ in eq. (1) is now given by $\xi ' = \ell n \
x_1/x$. \par

The crucial point of the present work is the following~: using the reggeon
factorization property
in eq. (2) it is possible to relate the Pomeron structure function $F_P$ to the
proton structure
function $F_2^p$ (or more precisely to that of the deuteron $F_2^d$, since the
isospin of the
Pomeron is equal to zero). Furthermore, the parameters in $F_P$ are entirely
given in terms of
those in $F_2^d$, plus a few reggeon couplings which can be obtained from soft
hadronic diffraction
in the framework of conventional Regge theory. \par

\vskip 5 mm
\noindent {\bf \underbar{The model}.} \ In a previous paper$^{[6]}$, we have
introduced the following
parametrization of the proton (and deuteron) structure function $F_2^p(F_2^d)$
at moderate values
of $Q^2$, based on Regge theory~:
\vfill \supereject
$$F_2(x, Q^2) = A \ x^{- \Delta(Q^2)}(1 - x)^{n(Q^2)+4} \left ( {Q^2 \over Q^2
+ a} \right )^{1 +
\Delta(Q^2)}$$
$$+ B \  x^{1 - \alpha_R} (1 - x)^{n(Q^2)} \left ( {Q^2 \over Q^2 +
b} \right )^{\alpha_R} \eqno(5)$$
\vskip 3 mm
\noindent with

$$\Delta(Q^2) = \Delta_0 \left ( 1 + {2 Q^2 \over Q^2 + d} \right ) \quad ,
\qquad n(Q^2) = {3 \over
2} \left ( 1 + {Q^2 \over Q^2 + c} \right )$$
\vskip 3 mm

\noindent where $1 + \Delta(Q^2)$ is the Pomeron intercept and $\alpha_R$ that
of the
secondary reggeon. The only secondary trajectory that contributes to $F_2^d$ is
the $f$ one. By
comparing the Pomeron structure function (Fig. 2) with that of the nucleon
(Fig. 3), and using
factorization, we see that the Pomeron structure function $F_P$ is identical to
$F_2^d$, given
by eq. (5), except for the following changes in its parameters

$$F_P(x_1, Q^2, t) = F_2^d \left ( x_1 , Q^2 ; A \to eA, B \to f B, n(Q^2) \to
n(Q^2) -
2 \right ) \eqno(6)$$
\vskip 3 mm
\noindent and where

$$e = {r_{PP}^P(t) \over g_{pp}^P(0)} \qquad , \quad f = {r_{PP}^f(t) \over
g_{pp}^f(0)} \ \ \ . $$
\vskip 3 mm

\noindent Note that the $t$-dependence of $F_P$ is entirely due to that of the
triple reggeon
couplings $r(t)$. Comparison with experiment shows that the $t$-dependence of
$r_{PP}^P$ and
$r_{PP}^f$ is practically the same and very weak. We have incorporated it in
the $t$-dependence of
$g_{pp}^P(t) = g_{pp}^P(0) \ exp \ (Ct)$, with $C = 2.2 \ {{\rm
GeV}^{-2}}^{[3]}$. In this way
$e$, $f$ and $F_P$ become independent of $t$. All the parameters in $F_2^d$ are
given in ref. [6].
In particular $B = B_u + B_d = 1.2$. \par
The parameters $e$ and $f$ in $F_P$ are obtained from
conventional triple reggeon fits to high mass single diffraction dissociation
for soft hadronic
processes. The most important point in relating soft and hard diffraction
dissociation is the
following. As discussed in ref. [6], absorptive (or sha\-do\-wing) corrections
due to rescattering
are very small at large $Q^2$ but may be quite large at $Q^2 = 0$. This is
particularly true for
diffractive processes. Indeed, it has been shown in ref. [16] [17] that
absorption corrections to
the proton diffractive cross-section are quite large~: they reduce the value of
the unabsorbed
cross-section by a factor of 3 to 4. (In the case of pions and photons, vector
mesons, the effect is
somewhat smaller). Since absorptive corrections decrease very rapidly when
$Q^2$ increases,
it is clear that, in the above considerations, valid at moderate and large
values of $Q^2$, we deal
with unabsorbed cross-sections. The values obtained$^{[17]}$ in this way are~:
$e = 3f = 0.1$. It
should be stressed, however, that, while the value of $e$ is rather well
determined, there are
uncertainties in the determination of $f$. In conventional triple reggeon
fits$^{[3]}$ with the
Pomeron intercept equal to 1, one obtains for the values of $e$ and $f$ in the
absorbed diffractive
cross-sections $e \simeq f$. With a Pomeron intercept above 1 (as the one in
the present paper),
the value of $e$ is practically unchanged while the value of $f$ decreases by a
factor of 2 to 3.
Precise data from HERA should allow a good determination of the parameter $f$.
\par

The Pomeron trajectory in the second factor of eq. (1) is parametrized as
$\alpha_P(t) = 1 +
\bar{\Delta} + \alpha 't$ with $\alpha ' = 0.25 \ {\rm GeV}^{-2}$. Here we use
the value of the
intercept at $Q^2 \leq  1 \ {\rm GeV}^2$. This is due to the fact that the high
virtuality of the
photon does not ``penetrate'' into the lower part of the diagram of Fig. 3.
Following refs. [16]
and [17] we use $\bar{\Delta} = 0.13$, which corresponds to the effective
Pomeron intercept
obtained without eikonal absorptive corrections. (In view of the smallness of
the triple reggeon
coupling $r$, the eikonal absorptive corrections in the lower part of the
diagram are expected to be
small). \par

The only remaining parameter is the proton-Pomeron coupling $g_{pp}^P(0)$ for
which we use
the value $(g_{pp}^P(0))^2 = 23$ mb$^{[16][17][18]}$. \par

Apart from the change in the parameters
resulting from Regge factorization, the Pomeron and proton (or deuteron)
structure functions also
differ in the $x \to 1$ behaviour. The Dual Parton Model arguments relevant for
$Q^2 = 0$
lead$^{[19]}$ to $n(0) = - 1/2$ (as compared to 3/2 for the proton), and
dimensional counting rules
relevant for $Q^2 \not= 0$ lead$^{[20]}$ to $n = 1$ (as compared to 3 for the
proton). This provides
the justification for the change in $n(Q^2)$ introduced in eq. (6). \par

As discussed in our previous paper$^{[6]}$, the parametrization in eqs. (5) and
(6) has to be
used at moderate $Q^2$ (up to $Q^2 \sim 5 \div 10 \ GeV^2$). At
larger values of $Q^2$, we use eq.~(6) as an initial condition for QCD
evolution, proceeding
as in ref. [6]. In doing so we determine $F_P$ at all values of $Q^2$. Note
that in
order to perform the QCD evolution we have to know the gluon distribution
function. In the proton
case, the normalization of this distribution was obtained$^{[6]}$ using the
energy-momentum
conservation sum rule. In the Pomeron case, $\sigma_{\gamma^{\ast} P}^{tot}$ is
not an ordinary
cross-section (see the discussion following eq. (1)), and therefore this sum
rule cannot be applied.
Fortunately, the normalization of the gluon structure function can be obtained
in this case using
Regge factorization, i.e. multiplying the gluon distribution in the proton,
given in [6], by the
factor $e = 0.1$. However, one expects that the gluon distribution in the
Pomeron is harder than
the one in the nucleon. In this case, the QCD evolution would be modified and
this would produce
some changes in our predictions (see below). \par

\vskip 5 mm
\noindent {\bf \underbar{Numerical results}.} \ In order to compare our
predictions with
preliminary HERA data we first compute the contribution to $F_2$ of the
diffractive events. This
is obtained from eq. (6) upon integration in $x_1$ and $t$. Using eqs. (1) to
(3) we have

$$F_2^{DD}(x, Q^2) = {\displaystyle{1.3} \over \displaystyle {16 \pi}}\int
{\displaystyle{ds_1} \over \displaystyle{s_1}} \int dt \left ( g_{pp}^P(t)
\right )^2 exp \left [ 2
\left ( \alpha_P(t) - 1 \right ) \xi ' \right ] F_P(x_1, Q^2, t) \ \ \ .
\eqno(7)$$
\vskip 3 mm

\noindent The integral $s_1$ runs from $0.4 \ {\rm GeV}^2$ to $(x_P^{max}/x -
1)Q^2$. The
dependence on the Bjorken-$x$ variable is due to the dependence of $\xi '$ on
$x$ and to the upper
limit of the integration on $s_1$. This value is determined using the cut $x_P
\leq 0.01$ present in
the HERA data. Note also that HERA data contain both single and double
diffraction. The latter is
estimated to be $\approx 30 \ \%$ of the former. This accounts for the factor
1.3 in eq. (7). Our
results are presented in Fig.~4 for the $x$ dependence at fixed value of $Q^2$,
and in Fig.~5
for the $Q^2$ dependence at fixed value of $x$, and compared with preliminary
data from the H1
collaboration. The agreement both in shape and absolute value is
satisfactory.\par

In Fig. 6, we show the theoretical $x_P$ dependence of $F^{DD}_2(x_1,x_P,Q^2)$
for different bins of $Q^2$ and $x_1$. It has to be noticed
that in eq. (1) the $x_P$ dependence is basically contained in the
factor $x_P^{-2\bar{\Delta}}$, the Pomeron structure function of eq. (6)
being independent of it. This explains why the $x_P$ dependence of
$F^{DD}_2(x_1,x_P,Q^2)$ appears to be the same for all bins when only
one Pomeron is exchanged.
\par

We also present in Fig. 7 our prediction for the Pomeron structure function
obtained from eq.
(6) at different values of $Q^2$. It will be important to check at HERA whether
there is an
approximate Bjorken scaling for the Pomeron structure function and to study
effects of scaling
violation due to QCD-evolution. As it was emphasized above these effects are
sensitive to the
behaviour of the gluon distribution in the Pomeron for $x_1 \sim 1$. This point
deserves further
study. \par

In Reggeon theory$^{[21]}$, shadowing corrections to the proton structure
function due to
double-Pomeron exchange are equal in magnitude and opposite in sign to the
diffractive
cross-section. Thus, we can reliably calculate the amount of shadowing in deep
inelastic scattering
at very small $x$, by using our model in the kinematical regions not yet
covered by
experiment. More precisely, we compute the ratio $F_P(x, Q^2)/F_2^p(x, Q^2)$
where
$F_2^p$ is the proton structure function computed in ref. [6] and $F_P$ is
given by eq. (7)
with the value $s_1^{max}$ obtained using $x_P \leq 0.1$, and we show
the the theoretical result in Fig. 8. In the
region of large $Q^2$ and in the interval $10^{-4} < x < 10^{-3}$ the shadowing
effects
are rather small, ($16 \div 18 \ \%$ at $Q^2 = 15 \ {\rm GeV}^2$ and $
12 \div 14 \ \%$ at $Q^2 = 30 \ {\rm GeV}^2$, at $x = 10^{-4}$).
Therefore, at HERA one does study the properties of the unabsorbed Pomeron.
\par

Before concluding we would like to discuss the possibility of using the above
formulae to
describe diffractive production with real photons ($Q^2 = 0$). Although this is
possible, there
is, however, a subtility which is precisely related to the fact (already
discussed above)
that, for diffractive processes, absorptive or shadowing corrections due to
rescattering are very
small at large $Q^2$ but are quite large at $Q^2 = 0$. Thus, when we put in our
formulae
$Q^2 = 0$, we obtain the \underbar{unabsorbed} values of diffractive
cross-sections of real photons.
In order to compare with experiment, we have to correct the obtained
cross-section for absorption
(which amounts roughly to reduce by a factor of 2 to 3 the values resulting
from the above formulae).
In this way, the value of $R$ for real photons turns out to be $R \sim 0.30
\div 0.35$ - including
the contribution of low mass vector mesons ($\rho$, $\omega$, $\phi$). \par

In conclusion, using a parametrization of the structure functions$^{[6]}$ based
on Regge theory
together with Reggeon factorization, we have been able to describe the
properties of the large
rapidity gap events observed at HERA in terms of Regge parameters determined
from soft diffraction
in hadronic processes.

\vskip 5 mm
\noindent \underbar{\bf Acknowledgments} \par
It is a pleasure to thank E. Levin for interesting discussions, and J.
Phillips, from the H1
col\-la\-bo\-ra\-tion, for valuable information on the experimental data.

\vfill\eject
\centerline{\underbar{\bf References}} \bigskip
\item {1.} M. Derrick et al. (Zeus Collaboration), Phys. Lett. \underbar{B315},
481 (1993).
 \item{2.} J. Dainton (H1 Collaboration), RAL-94-012.
\item{} S. Levonian (H1), Mini-school on ``Diffraction at HERA'', Desy, May
1994. \par
\item {3.} A. Kaidalov, Phys. Rep. \underbar{50}, 157 (1979). \par
\item{4.} G. Alberi and G. Goggi, Phys. Rep. \underbar{74}, 1 (1981). \par
  \item {5.} K. Goulianos, Phys. Rep. \underbar{101}, 169 (1983). \par
\item {6.} A. Capella, A. Kaidalov, C. Merino and J. Tran Thanh Van, Prep.
LPTHE Orsay 94-34. \par
\item {7.} A. B. Kaidalov, K. A. Ter-Martirosyan, Nucl. Phys. \underbar{B75},
471 (1974).  \par
\item {8.} G. Ingelman and P. E. Schlein, Phys. Lett. \underbar{152B}, 256
(1985). \par
\item {9.} E. L. Berger et al., Nucl. Phys. \underbar{B286}, 704 (1987).  \par
\item {10.} A. Donnachie and P. V. Landshoff, Nucl. Phys. \underbar{B303}, 634
(1988). \par
\item {11.} M. G. Ryskin, Prep. DESY 90-050 (1990). \par
\item{12.} E. Levin and M. Wuesthoff, Photon diffraction dissociation in DIS,
Desy 92-166 and
Fermilab-pub-92/334-T, Nov. 92. \par
\item{13.} J. Bartels and M. Wuesthoff, Desy-94-016, Feb. 1994. \par
\item{14.} A. H. Mueller and B. P. Pubet, CU-TP-625, Feb. 94. \par
\item{15.} N. N. Nikolaev and B. G. Zakharov, Z. Phys. \underbar{C53}, 331
(1992).   \par
\item{16.} A. Capella, J. Kaplan and J. Tran Thanh Van, Nucl. Phys.
\underbar{B105} (1976) 333.
Ibid \underbar{B97} (1975) 493. \par
\item{17.} A. B. Kaidalov, L. A. Ponomarev and K. A. Ter-Martirosyan, Sov.
Journal of Nucl. Phys.
\underbar{44} (1986) 468. \par
 \item{18.} A. Donnachie and P. V. Landshoff, Phys. Lett.
\underbar{B296} (1992) 227. \par
\item {19.} A. Capella, U. Sukhatme, C-I. Tan and J. Tran Thanh
Van, Phys. Rep. \underbar{236} (1994) 225. \par
\vfill \supereject
\item{20.} S. Brodsky and G. Farrar, Phys. Rev. Lett. \underbar{31} (1973)
1153.
\item{} V. Matveev, R. Muradyan and A. Tavkhelidze, Lett. Nuovo Cim.
\underbar{7} (1973) 719. \par
\item {21.} V. N. Gribov, JETP \underbar{53}, 654 (1967). \par

\vfill \supereject
\centerline{\underbar{\bf Figures Captions}} \bigskip
{\parindent = 1 truecm
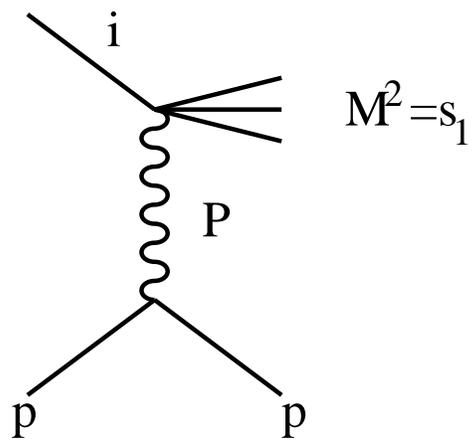
\item{\bf Fig. 1} Diffractive dissociation of particle $i$ in an $i$-$p$
collision. \par
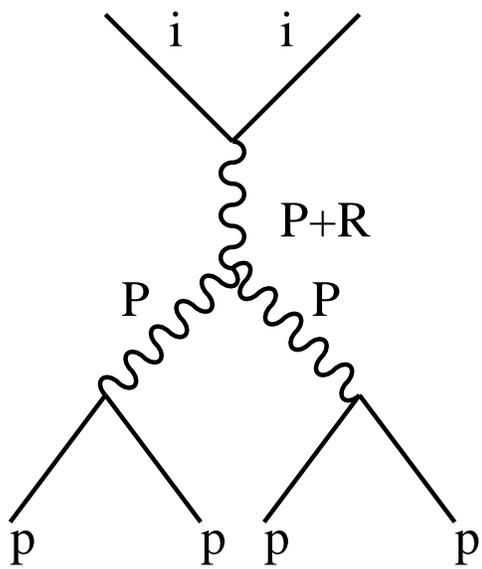
\item{\bf Fig. 2} Triple reggeon diagram obtained by squaring, in the sense of
unitarity, the
single diffraction dissociation diagram of Fig. 1.\par
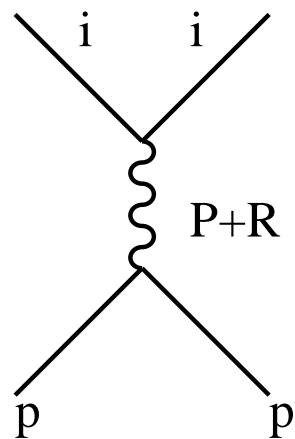
\item{\bf Fig. 3} Reggeon diagram for elastic $i$-$p$ scattering. When $i$ is a
virtual photon this
diagram corresponds to the proton structure function. \par
\item{\bf Fig. 4} Diffractive contribution to $F_2(x, Q^2)$ versus $x$, at
fixed $Q^2$. The
curves are obtained from eq. (7) with $x_P^{max} = 0.01$. The preliminary data
from the H1
collaboration, taken from the second item in ref. [2], have the same $x_P$ cut.
The upper (lower)
curves correspond to $e/f = 2$ ($e/f = 3$). \par
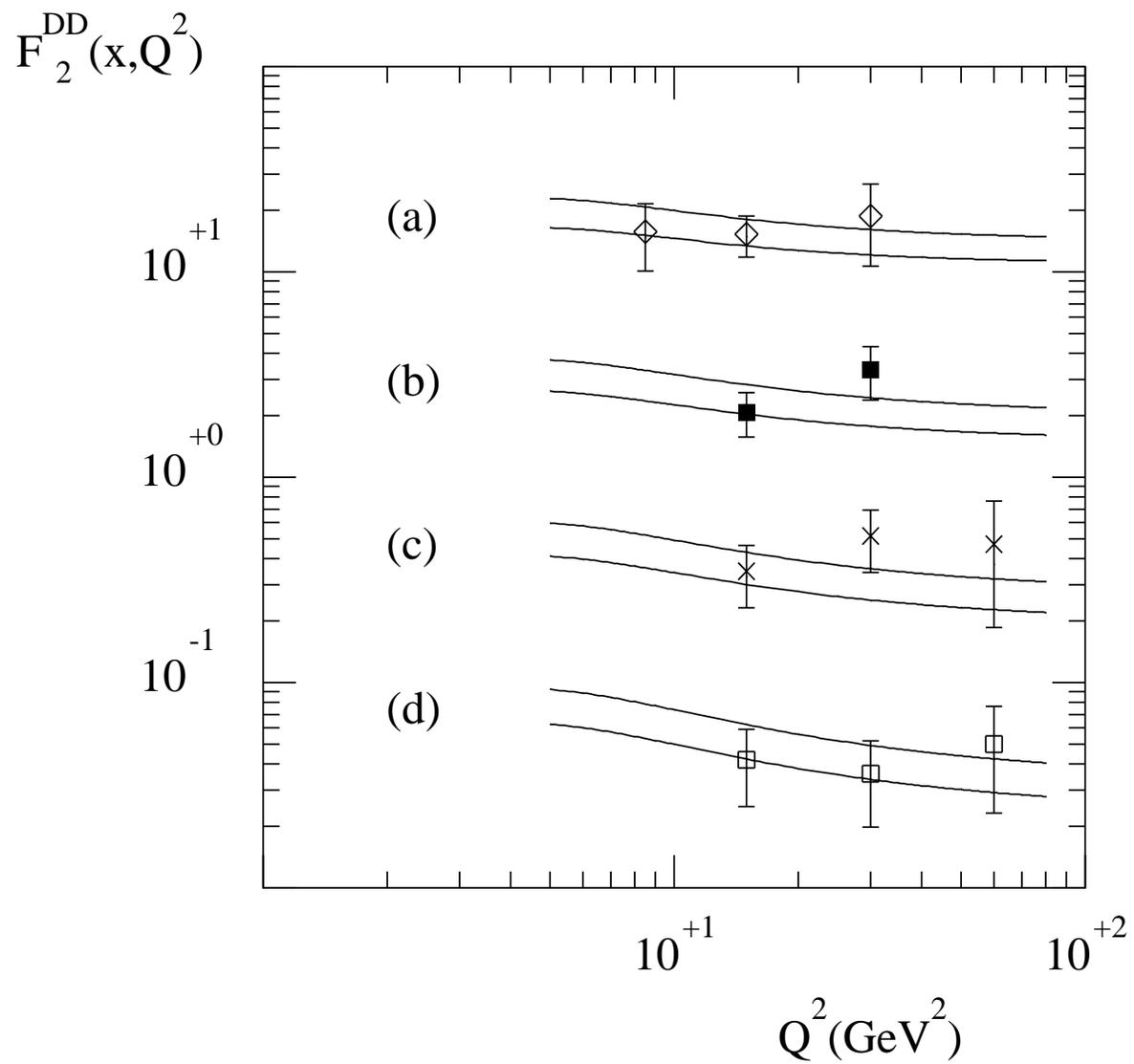
\item{\bf Fig. 5} The same as in Fig. 4 for $F_2^{DD}(x, Q^2)$ versus $Q^2$,
at fixed $x$. (a) x=0.00042(*125), (b) x=0.00075(*25), (c)
x=0.00133(*5), (d) x=0.00237(*1). \par
\item{\bf Fig. 6} $F^{DD}_2(x_1,x_P,Q^2)$ versus $x_P$ for different bins of
$Q^2$ and $x_1$. The upper (lower) lines correspond to $e/f = 2$ ($e/f
= 3$) in the case of single diffraction.\par
\item{\bf Fig. 7} The structure function of the Pomeron $F_P(x_1, Q^2)$
obtained from eq. (6). The curves in 1 (2) have been obtained by taking $e/f =
2$ ($e/f = 3$).\par
\item{\bf Fig. 8} Shadowing contribution versus $x$ for two different values
of $Q^2$. Full (dashed) lines have been obtained with $e/f = 2$
($e/f = 3$).\par}

\vfill\supereject
\centerline{  }
\vskip1.5cm
$$\hskip2.3cm\epsfbox{Fig94421}$$
\vskip1cm
$$\epsfbox{Fig94422}\hskip0.5cm\epsfbox{Fig94423}$$

\vfill\supereject
\centerline{  }
\vskip1.5cm
$$\epsfbox{Fig94424a}\epsfbox{Fig94424b}$$
$$\epsfbox{Fig94424c}\epsfbox{Fig94424d}$$
\centerline{Figure 4}

\vfill\supereject
\centerline{  }
\vskip3.cm
$$\epsfbox{Fig94425}$$

\vfill\supereject
\centerline{  }
\vskip1.5cm
$$\epsfbox{Fig94426a}\epsfbox{Fig94426b}$$
$$\epsfbox{Fig94426c}\epsfbox{Fig94426d}$$
\centerline{Figure 6}

\vfill\supereject
\centerline{  }
$$\epsfbox{Fig94427a}\epsfbox{Fig94427b}$$
\centerline{Figure 7}
\centerline{  }
$$\epsfbox{Fig94428}$$
\centerline{Figure 8}

\bye